\documentclass[conference]{IEEEtran}
\IEEEoverridecommandlockouts
\usepackage{cite}
\usepackage{amsmath,amssymb,amsfonts}
\usepackage{algorithmic}
\usepackage{graphicx}
\usepackage{textcomp}
\usepackage{hyperref}
\usepackage{slashbox}
\usepackage{booktabs}
\usepackage{multirow}
\usepackage{amsmath}
\usepackage{balance}
\usepackage{subfig}
\usepackage{float}
\hypersetup{breaklinks=true}

\usepackage[table]{xcolor}

\def\BibTeX{{\rm B\kern-.05em{\sc i\kern-.025em b}\kern-.08em
    T\kern-.1667em\lower.7ex\hbox{E}\kern-.125emX}}
\begin{document}

\title{Deep Learning on Chest X-ray Images\\ to Detect and Evaluate Pneumonia Cases \\at the Era of COVID-19}

\linespread{0.8}

\author{\IEEEauthorblockN{Karim Hammoudi$^{1,2,*}$}
\IEEEauthorblockA{\textit{$^{1}$Universit\'e de Haute-Alsace} \\
\textit{Department of Computer Science, IRIMAS}\\
F-68100 Mulhouse, France \\
\textit{$^{2}$Universit\'e de Strasbourg}\\
$^{*}$Corresponding author\\
karim.hammoudi@uha.fr}
\and
\IEEEauthorblockN{Halim Benhabiles$^{3}$}
\IEEEauthorblockA{\textit{$^{3}$UMR-8520--IEMN, Univ. Lille} \\
\textit{CNRS, YNCREA-ISEN}\\
Lille F-59000, France\\
halim.benhabiles@yncrea.fr}
\and
\IEEEauthorblockN{Mahmoud Melkemi$^{1,2}$}
\IEEEauthorblockA{\textit{$^{1}$Universit\'e de Haute-Alsace} \\
\textit{Department of Computer Science, IRIMAS}\\
F-68100 Mulhouse, France \\
\textit{$^{2}$Universit\'e de Strasbourg}\\
mahmoud.melkemi@uha.fr}
\and
\IEEEauthorblockN{Fadi Dornaika$^{4,5}$, Ignacio Arganda-Carreras$^{4,5}$}
\IEEEauthorblockA{\textit{$^{4}$University of the Basque Country} \\
\textit{Department of Computer Science}\\
\textit{\& Artificial Intelligence}\\
20018 San Sebasti\'an, Spain \\
\textit{$^{5}$IKERBASQUE, Basque Foundation for Science}\\ 
48011 Bilbao, Spain\\
fadi.dornaika@ehu.eus, ignacio.arganda@ehu.eus}
\and
\IEEEauthorblockN{Dominique Collard$^{6,7}$}
\IEEEauthorblockA{\textit{$^{6}$LIMMS/CNRS-IIS UMI 2820}\\
\textit{Institute of Industrial Science}\\
\textit{The University of Tokyo}\\
4-6-1 Komaba Meguro Ku\\
Tokyo 153-8505, Japan\\
\textit{$^{7}$CNRS/IIS/COL/Lille 1 SMMiL-E project}\\
\textit{CNRS D\'el\'egation Nord-Pas}\\
\textit{de Calais et Picardie}\\
2 rue des Canonniers\\
Lille, Cedex 59046, France\\
dominique.collard@yncrea.fr}
\and
\IEEEauthorblockN{Arnaud Scherpereel$^{8}$}
\IEEEauthorblockA{\textit{$^{8}$University of Lille}\\
\textit{CHU Lille, Inserm}\\
\textit{U1189 - ONCO-THAI}\\
F-59000 Lille, France\\
arnaud.scherpereel@chru-lille.fr}
}


\maketitle

\begin{abstract}
Coronavirus disease 2019 (COVID-19) is an infectious disease with 
first symptoms similar to the flu. COVID-19 appeared first in China and very quickly spreads to the rest of the world, causing then the 2019-20 coronavirus pandemic.
In many cases, this disease causes pneumonia. Since pulmonary infections can be observed through radiography images, this paper investigates deep learning methods for automatically analyzing query chest X-ray images with the hope to bring precision tools to health professionals towards screening the COVID-19 and diagnosing confirmed patients. In this context, training datasets, deep learning architectures and analysis strategies have been experimented from publicly open sets of chest X-ray images. Tailored deep learning models are proposed to detect pneumonia infection cases, notably viral cases. It is assumed that viral pneumonia cases detected during an epidemic COVID-19 context have a high probability to presume COVID-19 infections. Moreover, easy-to-apply health indicators are proposed for estimating infection status and predicting patient status from the detected pneumonia cases. Experimental results show possibilities of training deep learning models over publicly open sets of chest X-ray images towards screening viral pneumonia. Chest X-ray test images of COVID-19 infected patients are successfully diagnosed through detection models retained for their performances. The efficiency of proposed health indicators is highlighted through simulated scenarios of patients presenting infections and health problems by combining real and synthetic health data.    
\end{abstract}


\pagestyle{plain}
\section{Introduction}

COVID-19, initially named 2019-nCoV, appeared first in China and very quickly spreads to the rest of the world causing then the 2019-20 coronavirus pandemic.
To date (April 5th 2020), there have been 82,602 highly controversial confirmed cases in China, more than 500,000 confirmed cases in Europe and 1,226,644 confirmed cases all around the world\footnote{Online map of COVID-19 Global Cases by the CSSE center at Johns Hopkins University: \href{https://coronavirus.jhu.edu/map.html}{https://coronavirus.jhu.edu/map.html}}.

In many cases, this disease causes pneumonia. Characteristics of such an infection can be observed by radiologists. Also, deep learning methods can be helpful for operating deep analysis on query radiography images. Thanks to artificial intelligence, early stage and precision diagnosis can be done. 

In this pandemic, the effective screening of COVID-19 is an arduous task in practice. Standard screening test kits called Reverse Transcription-Polymerase Chain Reaction (RT-PCR) are often unavailable. Moreover, the RT-PCR test is highly sensitive. It was found that deep-based Computed Tomography (CT) images analysis could be more reliable than RT-PCR test in early-stage diagnostic \cite{doi:10.1148/radiol.2020200343,xu2020deep,Agarwal}. Notably, where RT-PCR test can turn out negative, deep CT image analysis can already predict true positives in certain cases. False negatives to RT-PCR test can lead to non-negligible propagation of this disease.

At this time, American College of Radiology recommendations for the use of chest radiography and CT for Suspected COVID-19 infection \cite{ACR} point that generally, the findings on chest imaging in COVID-19 are not specific, and overlap with other infections, including influenza, H1N1, SARS and MERS. Notably, being in the midst of the current flu season with a much higher prevalence of influenza in the U.S. than COVID-19, further limits the specificity of CT. Besides, the use of radiography equipment requires high disinfection needs after each use which can make massive tests laborious and time-consuming. In practice, for hygienic reasons, chest X-rays are often frontally taken with patients on a stretcher or bed, lying down or at best sitting. Such constraints often conduct to chest X-rays with poor quality and real issues in term of analysis. 

Nevertheless, CT images of lung and chest X-ray images offer additional data for screening COVID-19. Notably, AI technology is already deployed in China for radiography examination and radiomics-like analysis from CT images \cite{AIchina}. AI technology can also facilitate remote operations and help to face the lack of expert radiologists. At this date, many AI tools and radiography image datasets are private resources. The access to publicly open COVID-19-related sets of lung CT images towards conducting deep learning experiments is relatively limited. Some open access X-ray image sets of chest are publicly available. 

The goal of this paper is twofold: i) to present deep learning models tailored for detecting pneumonia infection cases such as viral cases towards screening COVID-19, ii) to propose easy-to-apply health indicators for evaluating detected pneumonia infection cases with an estimator of infection and predictions of patient status.
This case study is presented with the aim of supporting radiologists and other clinicians. In no case this preliminary study could be substituted to a medical advice.

The remaining of the paper is organized as follows. The next section gives an overview of related works. Section 3 presents some investigated deep learning based image detection architectures and analysis strategies. Section 4 shows a set of experiments to evaluate the performance of the considered architectures and section 5 concludes the paper.

\begin{figure*}[t]
    \centering{\includegraphics[width=\textwidth]{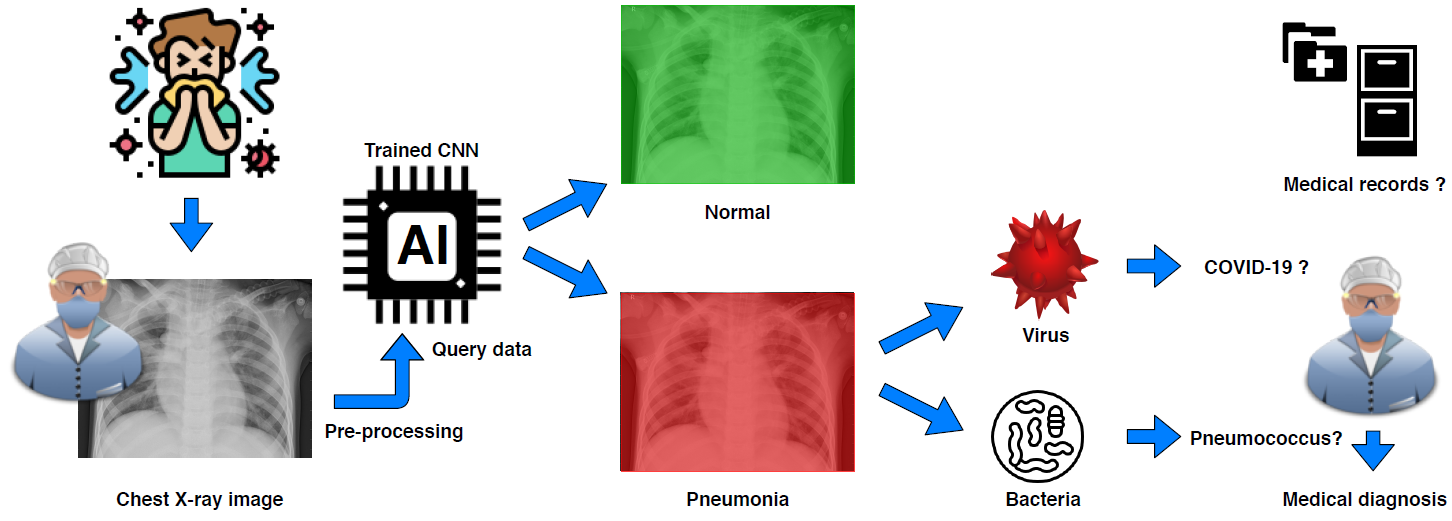}}
		\vskip -0.2cm
    \caption{Global workflow using deep learning for \textbf{automatic detection of infection towards supporting COVID-19 screening} from chest X-ray images. In a COVID-19 epidemic context, a detected viral pneumonia can particularly presume a COVID-19 infection.}
    \label{global_workflow}\vskip -0.2cm
\end{figure*}

\section{Related work}

Image analysis and machine learning techniques already have extensive applications in precision health. Currently, multiple and varied COVID-19-related studies are conducted in order to highlight proof of concepts and scientific truths about this misunderstood disease. Some image detection, evaluation and making-decision techniques related to COVID-19 and radiography examinations are described hereafter. 

In \cite{xu2020deep}, Xu et al. present a case study that deals with the COVID-19 screening from CT images. They mention that the COVID-19 manifests its own characteristics that differ from other types of viral pneumonia, such as Influenza-A viral pneumonia. The study aims to establish an early screening model for COVID-19 by automatically analyzing collected pulmonary CT images of COVID-19, Influenza-A viral pneumonia and healthy cases (618 transverse-section CT samples before data augmentation). The overall accuracy of the deep learning models were  86.7\% for these three groups.
First CT images were preprocessed for extracting effective pulmonary regions. Second, a 3D Convolutional Neural Network (CNN) model was used to segment multiple candidate regions (patches). In particular, the VNET \cite{MilletariNA16} based segmentation model VNET-IR-RPN \cite{wu2019deep} trained
for pulmonary tuberculosis purpose was used to separate candidate patches from viral pneumonia.
Third, a classification model (e.g. based on ResNet \cite{HeZRS15}) categorizes each patch amongst three types: COVID-19, Influenza-A-viral-pneumonia, and irrelevant-to-infection; and assign an infection probability. The overall analysis report for one CT sample was calculated using the Noisy-or Bayesian function.

In \cite{shan2020lung}, Shan et al. present a method to automatically segment and quantify infection in CT scans of COVID-19 patients. The collected dataset was composed of 549 CT images. To accelerate the manual delineation of CT images for training, a Human-In-The-Loop (HITL) strategy is adopted. A manual delineation step is operated by a medical staff for delimiting infected regions on original COVID-19 chest CT images. This permits to generate and to enrich a training set that is given as input to an artificial intelligence engine which operates to an automatic segmentation step over the COVID-19 infected region. The process loops these steps from this auto-contoured regions to assist radiologists for their annotation refinements. The proposed system yielded Dice similarity coefficients of approximately 91.6\% between automatic and manual segmentations, and a mean Percentage Of Infection (POI) estimation error of 0.3\% for the whole lung on the validation dataset. Moreover, compared with the cases of fully manual delineation that often requires 1 to 5 hours, the HITL strategy drastically reduces the delineation time to 4 minutes after 3 iterations of model updating.

The pulmonary infections can be more directly visible in CT images than in chest X-Ray images \cite{doi:10.1148/ryct.2020200034}. Nevertheless, detection of COVID-19 from chest X-ray images is also investigated since they represent widespread resources that are often analyzed upstream of CT scans.

In \cite{wang2020covidnet}, a deep learning architecture named COVID-Net is proposed for the detection of COVID-19 cases from chest radiography images. The authors notably exploits the \textit{Chest X-Ray Images (Pneumonia)} open dataset\footnotemark[1]\footnotetext[1]{\textit{Chest X-Ray Images (Pneumonia)} dataset: \href{https://www.kaggle.com/paultimothymooney/chest-xray-pneumonia}{https://www.kaggle.com/paultimothymooney/chest-xray-pneumonia}} and the \textit{COVID-19 Image Data Collection} open dataset \cite{cohen2020covid19}. Their derived chest radiography dataset named COVIDx is composed of 5941 posteroanterior chest radiography across 2839 patient cases. Their study targets the prediction of 4 image classes, namely normal, bacterial infection, non-COVID viral infection, and COVID-19 viral infection. In this sense, their dataset contains 1203 patient cases as normal, 931 patient cases with bacterial pneumonia, 660 patient  cases  with  non-COVID-19  viral  pneumonia and only 68 radiography images collected from 45 COVID-19 patient cases. The distribution of inter-class images amongst their training sets and amongst their test sets are highly unbalanced in view of the quantity of collected COVID-19 images. The authors exploit residual architecture design principles \cite{7780459} pointing that they show time and again to enable reliable neural network architectures that are easier to train to high performance. They leverage generative synthesis \cite{Wong} as the machine-driven design exploration strategy for generating the final COVID-Net network architecture that obtains a global test accuracy of 83.5\%. However, this accuracy has been obtained including a small data sample corresponding to only 10 COVID-19 cases.

\section{Proposed approach for pneumonia analysis}

\subsection{CNN-based detection and evaluation of infected patients}


In \cite{healthline}\cite{Landon}, authors emphasized that the COVID-19 is a viral disease and not a bacterial one. 
Respectively, an efficient classifier is designed to automatically detect if a query chest X-ray image is Normal, Bacterial or Viral by assuming that a COVID-19 infected patient, tested during an epidemic period, has a high probability to be a true positive when the classification output is Virus (see Fig. \ref{global_workflow}). Nethertheless, it is worth mentioning that a severe viral respiratory infection can lead to a secondary pneumonia of bacterial nature \cite{Hunter}.
For this reason, our classifier aims to be useful at early stage of COVID-19 pulmonary symptoms.

\subsubsection{Tailored CNN models}

A set of tailored models based on CNNs have been designed to take three set of image categories (e.g.; normal case, viral pneumonia case and bacterial case) as input and output the predicted probability for each of the categories. The trained models exploit the CNN backbones ResNet34, ResNet50 and DenseNet169 through the fastai library and a fully connected head, with a single hidden layer, as a classifier. 

Besides, a trained model exploits the CNN reference backbone VGG-19. In addition, a dual use model (Inception ResNetV2 - RNN) is prepared for i) characterizing  categories of input split images by getting a hidden layer output of a fin-tuned Inception ResNetV2 architecture, ii) predicting final categories of split images (image blocks) using a bidirectional Long Short-Term Memory (RNN-LSTM) architecture. For these last ones, a Keras and TensorFlow workflow is used. 

\begin{figure*}[!h]
    \centering{\includegraphics[width=\textwidth]{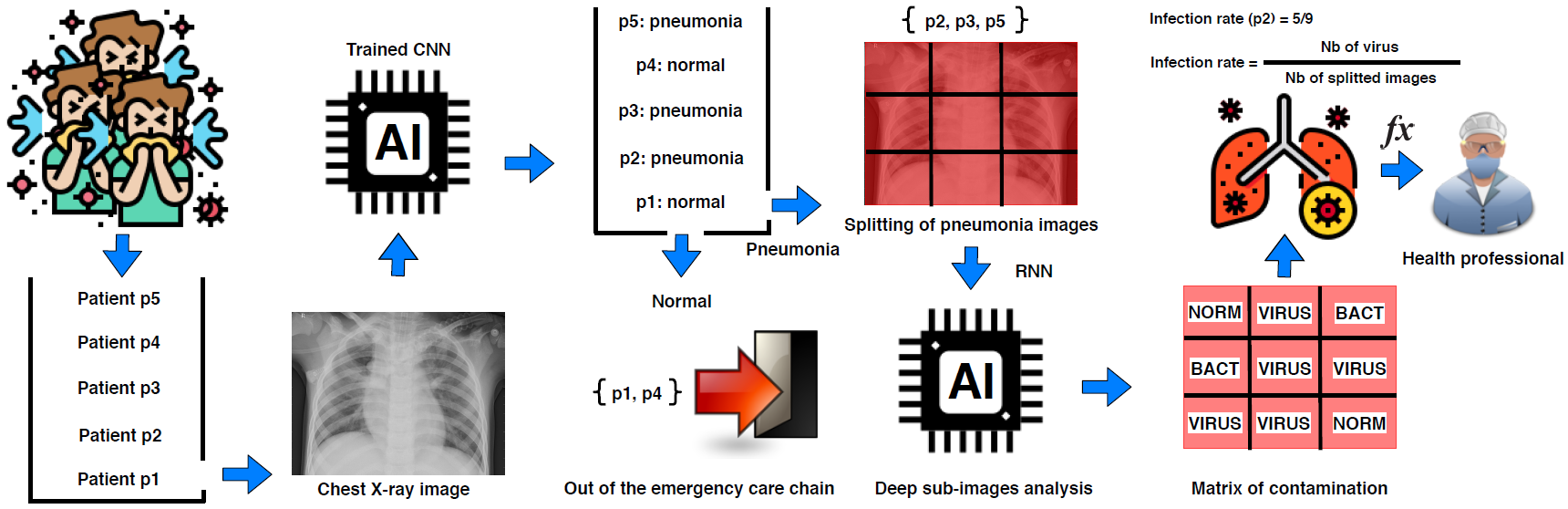}}
		\vskip -0.2cm
    \caption{Global workflow using deep learning based for \textbf{automatic estimation of a CNN-based infection rate indicator} from chest X-ray images.}
    \label{workflow_prioritization}\vskip -0.3cm
\end{figure*}

Specifically, the prediction stage of the dual-use model operates at a second level analysis of the data. A sequence of sub-images is first generated while entirely covering the images by directly positioning a regular grid onto the original query chest X-ray images. Precisely, the image is split into a set of image blocks that correspond to grid cells (see Fig. \ref{workflow_prioritization}). This operation enhances the size of the training set while limiting the loss of image details. This loss often occurs when the original images are resized for fitting inputs of standard deep learning architectures. Then, each image block is given as input to the RNN for providing a set of local predictions (matrix of contamination) towards estimating health indicators such as a CNN-based infection ratio (a use is described in section \ref{indicators}). The grid discretization should be tuned according to the obtained predictive performance of the considered architecture. 

\subsubsection{Data preparation and model inputs}

\begin{figure}[H]
  \begin{center}
    \subfloat[Normal.]{
      \includegraphics[width=1.9cm]{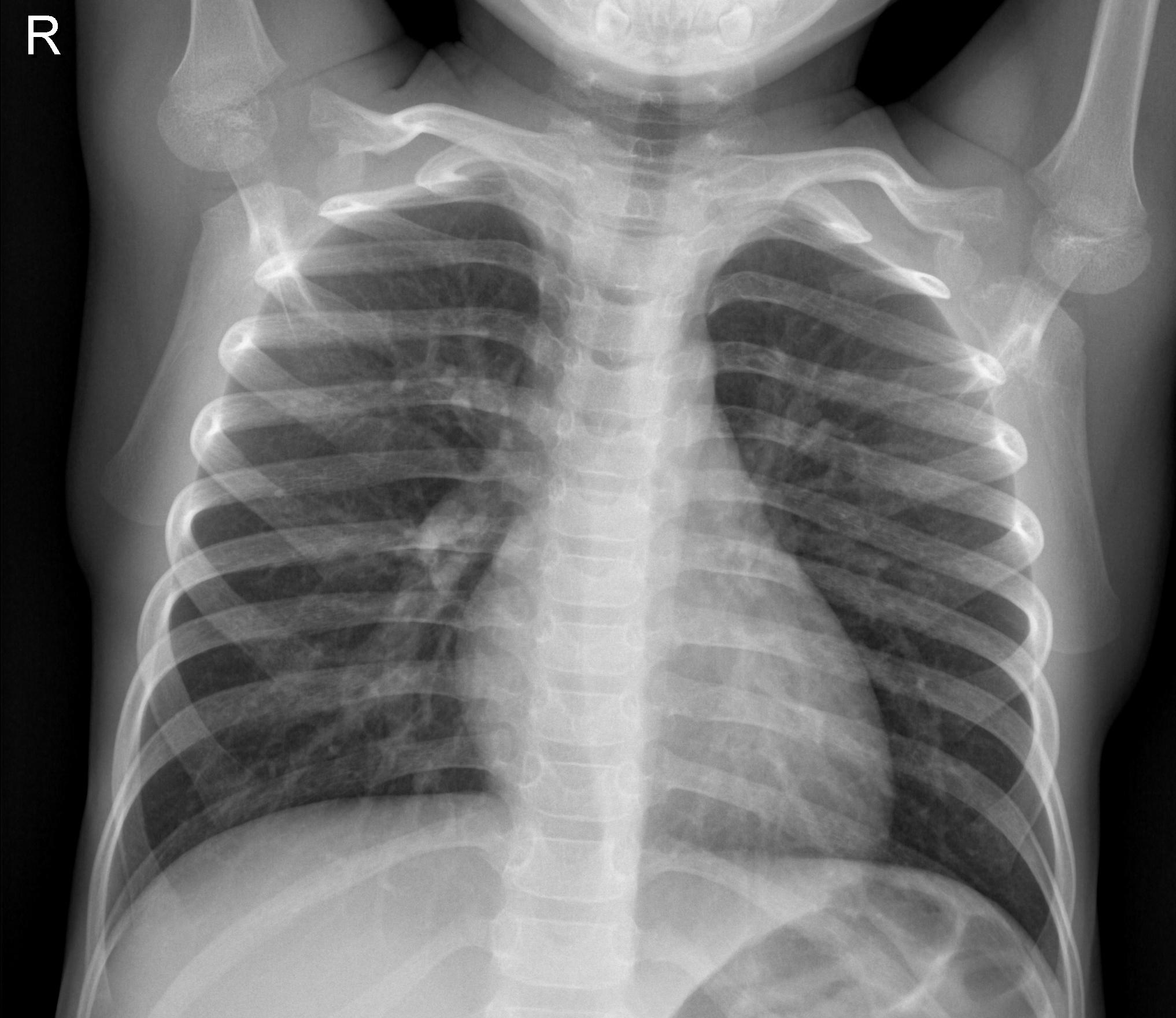}
      \label{norm}
                         }
    \subfloat[Bacteria.]{
      \includegraphics[width=1.9cm]{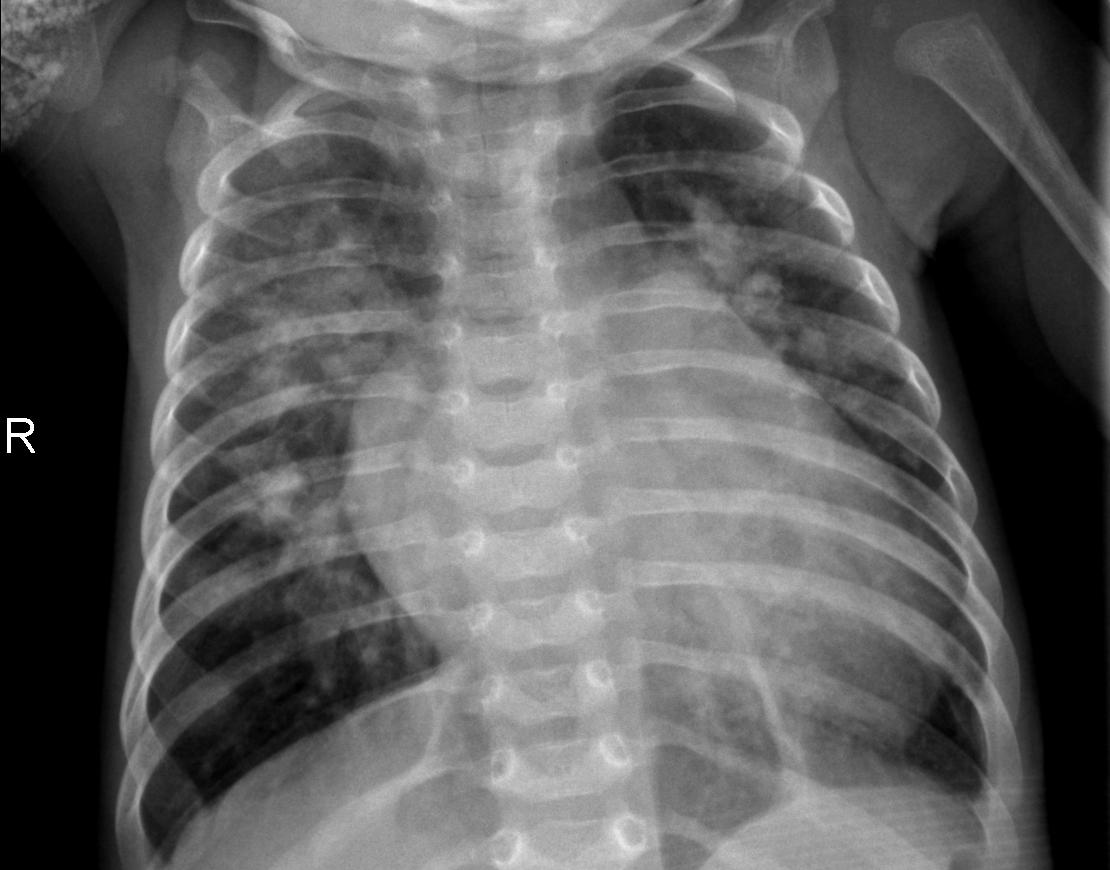}
      \label{bact}
                         }
    \subfloat[Virus.]{
      \includegraphics[width=1.9cm]{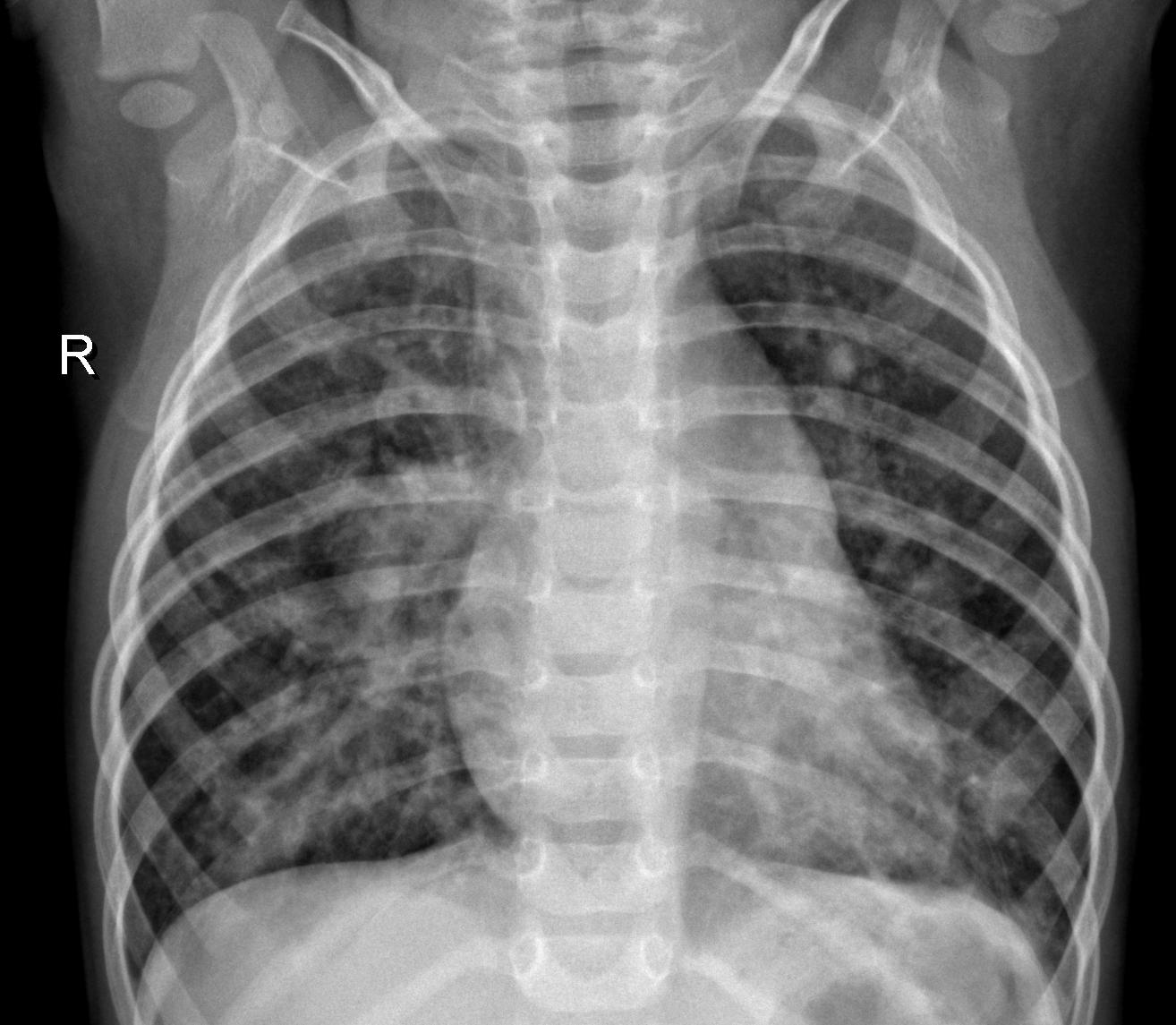}
      \label{virus}
                         }
    \subfloat[COVID-19.]{
      \includegraphics[width=1.9cm]{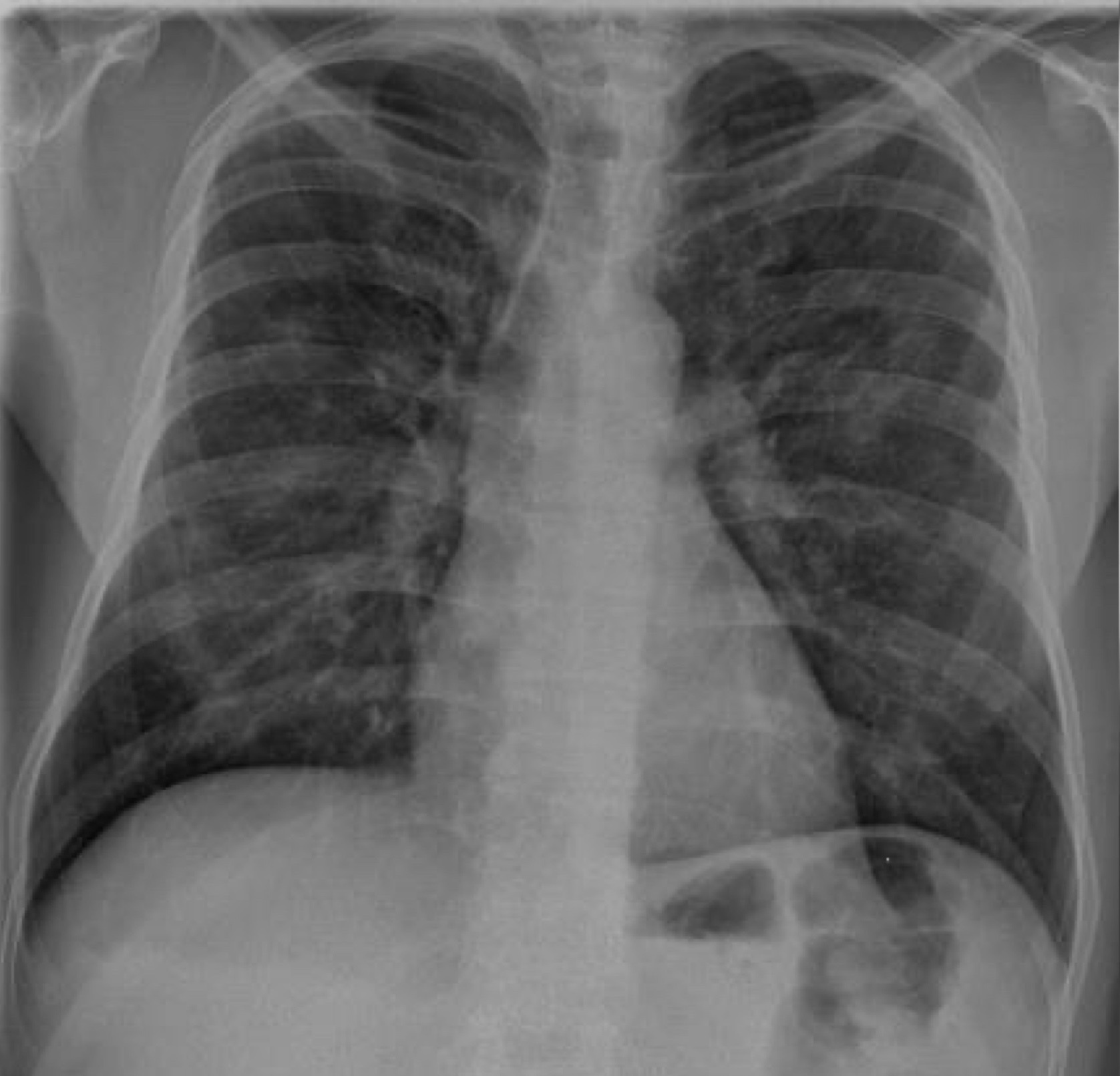}
      \label{covid}
                         }                                                                                                         
    \caption{Chest X-ray samples from the test datasets.}
    \label{data_sample}
  \end{center}
\end{figure}

For our experiments, we exploited Chest X-ray images from the \textit{Chest X-Ray Images (Pneumonia)} dataset\footnotemark[1].
This dataset is related to the paper \cite{datasetMooney} on the identification of medical diagnoses and treatable diseases by image-based deep learning. 
This dataset contains 5,863 children X-Ray images divided in two categories, namely Normal and Pneumonia. 
The Pneumonia category is composed of pneumonia images that are labeled either bacterial or viral (see illustrations in Fig. 6 of \cite{datasetMooney}).

The \textit{Chest X-Ray Images (Pneumonia)} dataset is reorganized into three classes; into normal, bacterial pneumonia and viral pneumonia (see samples in Fig \ref{norm}, Fig. \ref{bact} and Fig. \ref{virus}, respectively). Each training set contains 1345 images and each test set contains 148 images. Since this dataset was composed of pulmonary images having heterogeneous and large sizes; and to deal with reasonable computational times during the CNN training experiments, all the images were resized to a unique dimension and rescaled into smaller images (e.g.; size $310 \times 310$) to fit with standard inputs of tailored architectures. For the last tailored model using RNN, a preliminary split of the original image precedes the resizing step.  

Regarding the tests, we added a test set (blind test) that is composed of a single class containing 145 chest X-ray images of COVID-19 infected patients (see sample in Fig. \ref{covid}). This test set has been constituted by filtering the heterogeneous \textit{COVID-19 Image Data Collection} dataset \cite{cohen2020covid19}; folder containing a mix of CT and X-ray images with a variety of infection types. At this time, we consider that this quantity of available COVID-19 is still too limited for building a reliable detector that can discern between Non-COVID-19 viral pneumonia and COVID-19 viral pneumonia. The 145 chest X-ray images are specifically used as a test set towards ideally detecting them as viral pneumonia. As previously mentioned, we assume that a COVID-19 infected patient, tested during an epidemic period, has a high probability to be a true positive when the classification output is viral pneumonia (Fig. \ref{global_workflow}).

\subsection{Estimation of CNN-based health indicators}\label{indicators}

Based on statistical tools (logistic regression and statistical tests) and realistic data, studies on COVID-19-related death risk factor have been proposed in \cite{Fei2020,Caramelo2020}.
In this paper, we sketch a simple measure to provide to health professionals an estimator for evaluating the chance of a patient to survive COVID-19 considering risk factors; namely age, comorbidity and the infection rate indicator (Fig.\ref{workflow_prioritization}). For each risk factor, we associate a score which represents a penalty (a large value decreases the chance of a patient to escape fatality). The proposed measure $F$ is the addition of scores divided by a critical threshold $\mathcal{T}$. Beyond $\mathcal{T}$, there is no chance to be recovered. Formally, $F$ is expressed as follows:
    $$F = (\mathcal{S}_1 + \mathcal{S}_2 + \mathcal{S}_3)/\mathcal{T}$$
{\small where $\mathcal{S}_1$ measures the risk due to the patient's age, $\mathcal{S}_2$ measures the risk related to the CNN-derived infection rate measured from the X-ray chest image of a patient, $\mathcal{S}_3$ measures the risk associated with comorbidities (additional diseases) of a patient that can lead to the development of complications.}

More precisely, let us give an example to concretely compute the measure $F$. First, we point out that the scoring system used hereafter must be adjusted by health professionals to match with reality.

In this example, we use the values of fatality risk-ratio during COVID-19 epidemic in Hubei, China \cite{Riou2020}. Proportionally to these values, we define penalty scores ($\mathcal{S}_1$), see Table \ref{table0}.

\begin{table}[t]
 \caption{\textit{Scores related to age. The source of the ``Fatality risk ratio'' is \cite{Riou2020}.}}
  \centering {\scriptsize
  \begin{tabular}{|c||c|c|c|}
  \toprule
    {\textbf{Age}} & \textbf{Fatality risk ratio (Hubei, China)}  & \textbf{Associated score ($\mathcal{S}_1$)}  \\
   \hline\rowcolor{lightgray}
	 \centering $\geq 80$ & $18\%$ & \textbf{100} \\
  \hline
	 \centering 70-79 & 9.8\% & \textbf{54.4} \\
   \hline\rowcolor{lightgray}
   \centering 60-69 & 4.6\% & \textbf{25.5} \\
   \hline
   \centering 50-59 & 1.3\% & \textbf{7.2} \\
   \hline\rowcolor{lightgray}
   \centering 40-49 & 0.4\% & \textbf{2.2} \\
   \hline
   \centering 30-39 & 0.18\% & \textbf{1} \\
\hline\rowcolor{lightgray}
   \centering 20-29 & 0.09\% & \textbf{0.5} \\
   \hline
   \centering 10-19 & 0.02\% & \textbf{0.1}\\
   \hline\rowcolor{lightgray}
   \centering 0-9 & 0.01 & \textbf{0.05}\\
   \bottomrule
  \end{tabular}
}
\label{table0}
\end{table}

Then, we define the scores ($\mathcal{S}_2$) related to the {\it infection rate of X-ray image}. $\mathcal{S}_2$ can be the probability of the concerned class that is directly provided by the used CNN. A more refined formulation of $\mathcal{S}_2$ is proposed to scrutinize the X-ray image. Each image is divided in $n$ sub-images where $n=9$ (value for which our RNN effectively performs). After analysis, each sub-image will be in status virus, bacteria or normal (see Fig.\ref{workflow_prioritization}). A score equal to 100 is assigned to the X-ray image when the $n$ sub-images are infected by a virus. Proportionally, the infection rate of the image is:

$\mathcal{S}_2 = (100/n)\times N$, where $N$ is the number of the virus infected sub-images.

The third risk factor ($\mathcal{S}_3$) is related to diseases of a patient in addition to COVID-19. We associate a penalty 100 to serious diseases, as complicated hypertension, coronary artery disease, heart surgery, chronic renal failure on dialysis, cancer on treatment. The health professional can define a list of moderated diseases which may complicate the patient's recovering. He
may assign to a moderated disease score a penalty value
proportional to the score of serious diseases. In this article, we attribute the value 10 to one moderated disease.
In this example, we set the critical threshold $\mathcal{T}$ to 200, then beyond this limit, one cannot escape fatality. The measure F can be used as follows. If $ F \geq 1 $ then the hope to escape fatality is null.  Varying $F$ from 0 to 1, the patient gradually moves away from the hope of recovering.

The value 200 assigned to $\mathcal{T}$ is obtained by taking reference a patient having a serious additional disease and aged over 80 years. We assume that such a patient cannot fight against COVID-19. Therefore, a person having COVID-19 which reaches the score 200 cumulates too many factors to overcome the illness. 
 
Finally, we point out that the measure $F$ can also be extended by slightly modifying the term $\mathcal{S}_2$ for taking into account time factor. Indeed, a serie of X-ray images is observed for each COVID-19 inpatient pointed in \cite{doi:10.1148/ryct.2020200034}\cite{Zhang}\cite{ChenLancet}. Accordingly, $\mathcal{S}_2$ can be measure at two timely spaced X-rays $t_1$ and $t_2$ to take into account the time kinetics of symptom onset and disease progression for the infected patient. For this latter case, the infection rate can be redefined as follows:
\[
  \mathcal{S}_2=
\begin{cases}
    f(t_2) + malus,& \text{if } f(t_2) > f(t_1)+\delta~(aggravation)\\
   f(t_2),         & \text{if } |f(t_2) - f(t_1)| \leq \delta ~(stability)\\
   f(t_2) + bonus, & \text{otherwise}~(remission)
\end{cases}
\]
{\small where  $f(t) = (100/n)\times N(t)$ is the infection rate of the chest X-ray image captured at a moment $t$, $N(t)$ is the number of detected viral sub-images at a moment $t$, $t_1$ is inferior to $t_2$, $\delta$ is threshold fixed to 20, bonus and malus are a gain and a penalty fixed to $(-20\times f(t_2))/100$ and $(20\times f(t_2))/100$, respectively.}

\begin{table*}[!t]
\centering
\caption{Comparison of average accuracies obtained on classification using some tailored CNN-based architectures.}
\begin{tabular}{|l||*{5}{c|}}
\cline{2-6}
\multicolumn{1}{c|}{} & \multicolumn{5}{c|}{\textbf{Average accuracy (\%)}}\\
\hline
\backslashbox{\textbf{Data}}{\textbf{Network}}
												&\textbf{ResNet34}&\textbf{ResNet50}& \textbf{DenseNet169} 	&\textbf{VGG-19} 	& \textbf{Inception ResNetV2 \textbf{\& RNN}}\\
\hline\hline
\textbf{Image size} 								&(310x310),(310x273)&(310x310),(310x273)&(310x310),(310x273)&(224x224)				& (300x300)				\\\hline\rowcolor{lightgray}
\textbf{Raw resizing} 							&  93.69  				& 93.47 					& 91.89									&	82.66 					&   - \\\hline
\textbf{Original ratio with padding} 							&  90.54		  		& 93.92						& \textbf{95.72}				& -								& 	-		\\\hline\rowcolor{lightgray}
\textbf{Split in 9 \& raw resizing} & - 							& - 							& - 										&									& \textbf{80.40} \\\hline
\end{tabular}
\label{tab1}
\end{table*}

\begin{table*}[!t]
\centering
\caption{Classification performance obtained by testing our best trained architectures with two query image sets. The first row reflects classification results of \cite{datasetMooney}.}
\begin{tabular}{|l||*{4}{c|}}
\cline{2-5}
\multicolumn{1}{c|}{} & \multicolumn{4}{c|}{\textbf{Accuracy (\%)}}\\\hline
\backslashbox{\textbf{Query test set}}{\textbf{Class ouput}} &\textbf{Bacteria}&\textbf{Virus}&\textbf{Normal}&\textbf{Pneumonia}\\
\hline
\hline\rowcolor{lightgray}
\textbf{Dataset \textit{Chest X-Ray Images (Pneumonia)}\footnotemark[1] by \cite{datasetMooney} (2-stage binary training outputs)} &\multicolumn{2}{c|}{90.7 (Bact. vs Vir.)}&\multicolumn{2}{c|}{\textbf{92.8} (Norm. vs Pneu.)}\\
\hline
\hline
\textbf{3-class dataset \textit{Chest X-Ray Images (Pneumonia)}\footnotemark[1] by DenseNet169} &	\textbf{97.97} & \textbf{96.62}	&	\textbf{92.57} &-\\\hline\rowcolor{lightgray}
\textbf{3-class dataset \textit{Chest X-Ray Images (Pneumonia)}\footnotemark[1] by VGG-19}  &87.84&81.08&79.05&-\\\hline
\textbf{3-class dataset \textit{Chest X-Ray Images (Pneumonia)}\footnotemark[1] by Inception ResNetV2 \& RNN (majority voting)}  &86.49&84.46 &63.52 & -\\\hline\rowcolor{lightgray}
\textbf{3-class dataset \textit{Chest X-Ray Images (Pneumonia)}\footnotemark[1] by Inception ResNetV2 \& RNN (by default)} & 90.54& 83.78 & 66.89& -\\
\hline
\multicolumn{1}{c|}{} & \multicolumn{4}{c|}{\textbf{Sensitivity (\%)}}\\\hline
\backslashbox{\textbf{Query test set}}{\textbf{Class ouput}} &\multicolumn{2}{c|}{\textbf{Virus}}&\multicolumn{2}{c|}{\textbf{Pneumonia}}\\\hline\rowcolor{lightgray}
\textbf{1-class blind set of varied COVID-19 images by DenseNet169} &\multicolumn{2}{c|}{45.51}&\multicolumn{2}{c|}{88.27}\\\hline
\textbf{1-class blind set of varied COVID-19 images by Inception ResNetV2 \& RNN (majority voting)}  &\multicolumn{2}{c|}{\textbf{60.64}}&\multicolumn{2}{c|}{95.12}\\\hline\rowcolor{lightgray}
\textbf{1-class blind set of varied COVID-19 images by Inception ResNetV2 \& RNN (by default)} &\multicolumn{2}{c|}{51.72}&\multicolumn{2}{c|}{\textbf{99.3}}\\\hline
\end{tabular}
\label{tab2}
\end{table*}

\section{Experimental study}

\subsection{Performance of tailored CNN models}


\begin{table}[H]
\centering
\caption{Confusion matrix.}
\begin{tabular}{|l||*{3}{c|}}
\hline
\backslashbox{\textbf{Actual}}{\textbf{Predicted}}
                                                                                                                                                                                 &\textbf{Bacteria}&\textbf{Normal}& \textbf{Virus}\\\rowcolor{lightgray}
\hline
\textbf{Bacteria}             &\textbf{145} & 1 & 2 \\\hline
\textbf{Normal}                                                                                                       &  0                      & \textbf{137} & 11                                                                                                                                  \\\hline\rowcolor{lightgray}
\textbf{Virus} &  4                         & 1                                                                   & \textbf{143}   \\\hline
\end{tabular}
\label{confusion_matrix}
\end{table}


Table \ref{tab1} presents performance for classification of normal and infection cases by using tailored CNN-based architectures. 
The DenseNet169 architecture has reached best performance with an average classification accuracy of 95.72\% from the \textit{Chest X-Ray Images (Pneumonia)} dataset\footnotemark[1]. The classification accuracies are 97.97\%, 96.62\% and 92.57\% for the class bacterial, virus and normal, respectively (see Table \ref{tab2}). The associated confusion matrix is shown in Table \ref{confusion_matrix}. The performance of our DenseNet model is competitive with performances obtained by \cite{datasetMooney} in average classification accuracy of bacterial and viral cases 90.7\%.

However, our RNN-based architecture is particularly sensitive to pneumonia cases with the blind COVID-19 test set since it detects pneumonia for 99.3\% using default setting. Also, it promisingly detects viral infection for 60.64\% considering majority voting in sequences. 
We stress the fact that the 145 COVID-19 images which have been extracted from \cite{cohen2020covid19} are highly heterogeneous. Notably, these extracted images come from at least 24 different hospitals over the world. The RNN output results show a particularly robust pneumonia detection of COVID-infected patients and satisfying viral detection in view of the diversity of exploited radiography sources. 

In \cite{cohen2020covid19}, a histogram shows that significant image quantity has been acquired during the first week of the start of symptoms or hospitalization.
Since the quasi-totality of pneumonia are detected, our models should be able to operate at an early detection stage. 

Also, a histogram shows a significant image distribution in term of age in between 20 and 80 years old. Since the \textit{Chest X-Ray Images (Pneumonia)} dataset\footnotemark[1] is principally collected from children (5,232 chest X-ray images) and since the quasi-totality of pneumonia are detected, models trained on children chest X-ray image database may be relevant for detecting pneumonia from adult chest X-ray images.

\subsection{Projection with the CNN-based health indicators}

Table \ref{tab3} shows the RNN-derived infection rates \textbf{$\mathcal{S}_2$} estimated from real pairs of successive X-ray images for 5 COVID-19 infected patients.
Table \ref{tab4} details examples of $F$ calculated for 9 patients from synthetic data.

\begin{table}[t]
\centering\tiny
\caption{CNN-derived infection rates \textbf{$\mathcal{S}_2$} estimated from real pairs of successive X-ray images for 5 COVID-19 infected patients.}
\begin{tabular}{|p{1.5cm}|c|c|c|p{1.5cm}|c|}
\toprule
\centering\textbf{Image pairs} &\textbf{$f(t_1)$}&\textbf{Elapsed days}&\textbf{$f(t_2)$}&\centering\textbf{Observations}&\textbf{$\mathcal{S}_2$}\\
&&\textbf{($t2-t1$)}&&&\\
\hline
\hline \rowcolor{lightgray}
\textbf{(1,3), Fig. 5 of \cite{doi:10.1148/ryct.2020200034}} &\textbf{9/9}&\textbf{7}&\textbf{9/9}&\textbf{elderly man}&\textbf{$f(t_2)$}\\\rowcolor{lightgray}
 &&&&\textbf{``improvements''}&\\
\hline
\textbf{(1,2), Fig. 1 of \cite{Zhang}} &\textbf{1/9}&\textbf{5}&\textbf{2/9}&\textbf{67-y-old woman}&\textbf{$f(t_2)$}\\
 &&&&\textbf{``wires, attenuation''}&\\
\hline \rowcolor{lightgray}
\textbf{(1,3), Fig. 2 of \cite{Zhang}} &\textbf{5/9}&\textbf{8}&\textbf{5/9}&\textbf{36-y-old man}&\textbf{$f(t_2)$}\\\rowcolor{lightgray}
 &&&&\textbf{``death''}&\\
\hline
\textbf{(1,2), case 1, Fig. 5 of \cite{ChenLancet}} &\textbf{6/9}&\textbf{1}&\textbf{9/9}&\textbf{``worse status''}&\textbf{$f(t_2)$+malus}\\
\hline\rowcolor{lightgray}
\textbf{(1,2), case 2, Fig. 5 of \cite{ChenLancet}} &\textbf{7/9}&\textbf{4}&\textbf{8/9}&\textbf{``worse status''}&\textbf{$f(t_2)$}\\

\bottomrule
\end{tabular}
\label{tab3}
\end{table}

\begin{table}[t]
\centering\tiny
\caption{Examples of $F$ values for 9 patients from synthetic data.}
\begin{tabular}{|l|l|l|l|l|c|}
  \toprule
  \textbf{Patients} & \textbf{Aggravation factors: values} & \textbf{$\mathcal{S}_1$} & \textbf{$\mathcal{S}_2$} &\textbf{ $\mathcal{S}_3$} & \textbf{F}\\ 
	\hline
  \multirow{4}{*}{\textbf{Patient 1}} 
    & age: 82  & 100 & & &\\
    & nb. of infected sub-image: 3/9 & & 33.33 &  & \textbf{0.7166 (71.66\%)}\\
    & nb. of serious illness: 0 & & & 0 &\\
    & nb. of moderated illness: 1& & & 10 & \\ \hline
    
  \multirow{4}{*}{\textbf{Patient 2}} 
    & age: 50-59 & 7.2 & & &\\
    & nb. of infected sub-image: 4/9 & & 44.44 &  & \textbf{0.4582 (45.82 \%)} \\
    & nb. of serious illness: 0 & & & 0 &\\
    & nb. of moderated illness: 4& & & 40 & \\ \hline

\multirow{4}{*}{\textbf{Patient 3}} 
    & age: 70-79  & 54.4 & & &\\
    & nb. of infected sub-image: 1/9 & & 11.11 &  & \textbf{0.8275 (82.75\%)}\\
    & nb. of serious illness: 1 & & & 100 &\\
    & nb. of moderated illness: 0& & & 0 & \\ \hline

\multirow{4}{*}{\textbf{Patient 4}} 
    & age: 82  & 100 & & &\\
    & nb. of infected sub-image: 1/9 & & 11.11 &  & \textbf{1.05 ($>$ 100\%)}\\
    & nb. of serious illness: 1 & & & 100 &\\
    & nb. of moderated illness: 0& & & 0 & \\ \hline

\multirow{4}{*}{\textbf{Patient 5}} 
    & age: 30-39  & 1 & & &\\
    & nb. of infected sub-image: 5/9 & & 55.55 &  & \textbf{0.7827 (78.27\%)}\\
    & nb. of serious illness: 1 & & & 100 &\\
    & nb. of moderated illness: 0 & & & 0 & \\ \hline

\multirow{4}{*}{\textbf{Patient 6}} 
    & age: 10-19  & 0.1 & & &\\
    & nb. of infected sub-image: 7/9 & & 77.77 & & \textbf{0.8893 (88.93\%)}\\
    & nb. of serious illness: 1 & & & 100 &\\
    & nb. of moderated illness: 0 & & & 0 & \\ \hline

\multirow{4}{*}{\textbf{Patient 7}} 
    & age: 10-19  & 0.1 & & &\\
    & nb. of infected sub-image: 4/9 & & 44.44 & & \textbf{0.2227 (22.27\%)}\\
    & nb. of serious illness:0 & & & 0 &\\
    & nb. of moderated illness: 0 & & & 0 & \\ \hline

\multirow{4}{*}{\textbf{Patient 8}} 
    & age: 50-59  & 7.2 & & &\\
    & nb. of infected sub-image: 2/9 & & 22.22 & & \textbf{0.3471 (34.71\%)}\\
    & serious illness: 0 & & & 0 &\\
    & nb. of moderated illness: 4 & & & 40 & \\ \hline
 
\multirow{4}{*}{\textbf{Patient 9}} 
    & age: 40-49  & 2.2 & & &\\
    & nb. of infected sub-image: 3/9 & & 33.33 & & \textbf{0.2276 (22.76\%)}\\
    & nb. of serious illness: 0 & & & 0 &\\
    & nb. of moderated illness: 1 & & & 10 & \\ 
\bottomrule
\end{tabular}
\label{tab4}
\end{table}

\section{Conclusion}

A bench of deep learning tailored models have shown promising performances. Indeed, they have all exceeded 84\% of average accuracy on pneumonia detection cases for the Pneumonia reorganized dataset\footnotemark[1]. Hence, a patient that has a pneumonia during the epidemic context has a high probability to be detected by these models. In particular, the InceptionResNetV2 model has detected the minimum of false negatives to the pneumonia on the blind test set (0.7\%). 
Moreover, we have shown in our experiments that the transfer of knowledge from pediatric chest X-ray training towards infection screening of adults can be efficient. Additionally, an attempt based on realistic scenarios is done to provide easy-to-apply health indicators for evaluating infection rate and aggravation risk to the COVID-19 pneumonia.  
Future works may exploit our models to discern between COVID-19 viral and non-COVID-19 viral pneumonia once chest X-ray images of COVID-19 will be accessible in sufficient quantity. This should permit to specifically identify COVID-19 infected patients even in a non-epidemic context. Furthermore, reliability of proposed models must be cross-checked by RT-PCR tests and clinical tests before deployment.

\balance


\bibliographystyle{IEEEtran}
\bibliography{IEEEexample}

\end{document}